# An Application of Uncertain Reasoning to Requirements Engineering

**Philip S. Barry**
The MITRE Corporation
1820 Dolley Madison Blvd.
McLean, VA 22102-3481
pbarry@mitre.org

**Kathryn Blackmond Laskey**
Dept. of Systems Engineering
George Mason University
Fairfax, VA 22032-4444
klaskey@gmu.edu

## Abstract

This paper examines the use of Bayesian Networks to tackle one of the tougher problems in requirements engineering, translating user requirements into system requirements. The approach taken is to model domain knowledge as Bayesian Network fragments that are glued together to form a complete view of the domain specific system requirements. User requirements are introduced as evidence and the propagation of belief is used to determine what are the appropriate system requirements as indicated by user requirements. This concept has been demonstrated in the development of a system specification and the results are presented here.

## 1 Introduction

Requirements engineering consists of creating an agreement among developers, customers and users as to the intended functionality of a planned system, together with criteria for determining whether the completed system is acceptable. To reach this agreement, several distinct analysis steps must be taken. First, user requirements must be elicited. User requirements are necessary features, functions or attributes of a system that can be sensed from a position external to the system [DAV93]. Next, system requirements must be developed. System requirements are detailed specifications of the features or functions to be implemented, together with constraints on how they are to be implemented [SOM97]. Finally, both user requirements and system requirements are verified for completeness and consistency with each other and with user needs and domain constraints.

Because user requirements are often non-specific need statements and may engender a number of additional requirements, translating from user requirements to system requirements can be a difficult and error-prone process. Mannion and Keepence [MAN95] point out that the elicitation process frequently fails to produce a determination of the relative importance of requirements. Consequently, prioritization schemes during analysis may result in false emphasis due to the sheer number of requirements. Further, many requirements are dependent upon other requirements or may be at varying levels of abstraction. Such relationships are often not captured in either the user or the system requirements document.

The research discussed in this paper proposes that the relationships between system requirements in a domain can be modeled as a Bayesian Network. Specifically, fragments of Bayesian Networks are developed to model distinct aspects of the domain. For a particular requirements elicitation problem, an appropriate set of network fragments is combined to form a problem-specific Bayesian network. Evidence is introduced into the system in the form of user requirements. By observing the propagation of information through the network an assessment can be made as to what system requirements are implied by a given user requirement. This approach has the potential of more completely modeling the systemic implication of users needs leading to better system specifications and more accurate designs

## 2 A Conceptual Foundation for Use of Uncertain Reasoning in Requirements Modeling

To define a system in enough detail to begin design requires a coherent description of the system requirements. The first step in this exercise is to map the user requirements to system requirements. Frequently, this process can become quite involved as a given user requirement may engender a number of system requirements. Further, once a system requirement is implied, it may imply one or more additional system requirements. This expansion of system requirements contingent upon a user requirement, often referred to as allocation and flowdown [DOR97], is usually done using the requirements analyst's best judgement. Relevancy of system requirements to user requirements is often assessed heuristically or not at all.

When developing a new application in a new domain, the allocation and flowdown of requirements is a novel exercise and reuse of previous requirements definition exercises may not be possible. However, many software development projects are based upon a common technical base. Examples of such repeatable domains include groupware or help desk applications. We postulate that a mechanism to facilitate the reuse of the system requirements would lead to more rapid development of system requirement specifications as well as a more complete picture of the necessary system functionality.

To design such a mechanism, we first consider the question of how to represent the relationship of system requirements to each other and to the user requirements that engender them. To do this, we define an abstract structure of interrelated system requirements called a system requirement web (SRW). A SRW is a directed graph in which the nodes represent system requirements and the edges represent relationships between requirements that we call *weak implication*. We say that one node weakly implies another node if it is more likely that the requirement represented by the second node is needed if the first one is. User requirements can trigger a flow of weak implication within a SRW, providing a model for allocation and flowdown of requirements. Our research objective was to investigate the use of the SRW architecture as an aid to improve the efficiency and reliability of the process of generating a system requirements specification from user requirements.

Our initial attempt to implement the SRW concept was based on production rules. For example, to model the situation where requirement A is weakly implied by requirement B, a production rule was constructed that had an antecedent condition that B had been previously implied and a consequent that asserted requirement A as being newly implied. By creating a number of such production rules a directed graph of implication relationships was created. We implemented our production system in JESS [FRI98], which is a JAVA implementation of the CLIPS production system [GIA93].

Using production rules to model weak implication quickly ran into two difficulties. First, it was cumbersome to assess how strongly a given requirement was implied. This was important because strength of implication may have significant effect on design decisions. Second, the rule bases were difficult to generate and debug. In trial runs, we found that the system tended either to recommend far too few or far too many requirements.

This difficulty was addressed by introducing Bayesian networks to represent the weak implication of requirements by other requirements. The graphical structure of Bayes Nets provided a natural construction mechanism for SRWs with a clear path for diagrammatically creating the relationships between the requirements. Additionally, the quantitative nature of Bayes Nets offered excellent flexibility for modeling how strongly a requirement is implied. This modeling paradigm resulted in the calculation of a large number of conditional probabilities, but it was not considered to be a problem due to the availability of a number of off-the-shelf Bayesian Networks products.

Besides providing a useful visual metaphor that aided in the construction of SRWs, the basic concepts of SRWs were readily mapped into the Bayesian Network formalism. Specifically, SRW nodes were modeled as bi-valued random variables within the Bayesian Network, where the node (and the associated requirement) were either implied or not implied. The propagation of weak implication was modeled propagating evidence within the Bayesian Network. Representing weak implication using conditional probabilities provided a numeric assessment for the relevance of the system requirement as part of the Bayesian Network representation. In other words, the higher the probability that a system requirement is implied (P(implied)), the more likely that it is really appropriate. Additionally, the probability that the system requirement is not needed was easily represented by P(not implied), where P(not implied) = 1-P(implied).

We model a weak implication relationship from a set of requirements to another requirement by arcs in the Bayesian network from the nodes in the set to the implied requirement. We created generic combination operations for constructing conditional probability tables using "and" and "or" combinations. Probabilities near one in the conditional probability table indicate a that the system requirement is likely to be needed conditional on the configuration of implied values of the parent requirements. Probabilities near zero indicate the corresponding requirement is unlikely to be needed given the configuration of parent requirements.

As SRWs were constructed using Bayesian Networks, it became apparent that the behavior of large SRWs was often difficult to predict and test. We therefore broke up the larger SRWs into manageable chunks that corresponded to a natural decomposition of the domain of interest. By combining these fragments, various views into the domain could be created, emphasizing particular aspects that were important to the analyst.

The process for combining the SRW chunks was labeled gluing. A similar approach was taken by Laskey and Mahoney in their work on network fragments [LAS98]. Two SRW chunks are combined by unifying nodes common to the two chunks and assigning as parents to the resulting node the union of the parent nodes in the two input SRW chunks. The conditional probabilities are subsequently reallocated.

Two types of conditional probability allocations were used, a linear additive heuristic and the "noisy or". The linear additive heuristic is used in the instance where the nodes that weakly imply another SRW node from both sets of source SRWs work together, such that as more elements weakly imply the target SRW node the probability that the requirement is needed increases. The canonical rule for the "and" relationship within the SRW context is to first define the boundary probabilities for the node, basically the probability that the node is implied if all parents are implied and the probability the node is implied if none of the parents are implied. The conditional probabilities are then set using a linear distribution that treats each node equally. In the case study examined, most relationships between requirements could be modeled using the linear additive approach.

The "noisy or" model was used in the case where a child node has multiple parents and if one parent is weakly implied it will be sufficient to trigger the implication of the child. In this case, there is no synergistic effect between parents which differs from the linear additive approach. The "noisy or" also provides a effective model for the case where several distinct subsets of can imply the child requirement. A good discussion of the "noisy or" distribution is provided in [JEN96].

Pursuing the Bayesian Network implementation of SRWs engendered a requirement for numerous engineering approximations. Since no frequency information was available, there was no numeric way to assess prior and conditional probabilities. The use of traditional knowledge elicitation routes was quickly determined to be impractical due to the sheer number of assessments that would need to be made. This resulted in a situation where probabilities had to be assessed using general rules and tailoring when needed. As will be seen later in this paper, good results were achieved using this approach providing evidence for the proposition that in this problem the structure of the SRW mattered more than the specific probability assessments.

Bi-direction implication presented another challenge for the Bayesian Network approach. In theory SRW nodes can imply each other, and in effect create a cycle. From a requirements engineering standpoint this is perfectly logical. For example, a requirement for a database may imply a requirement for a database application and the requirement for the database application may imply the requirement for the database. However, when using Bayesian Networks a precedence relationship must be established and the direction of implication must be modeled in only one direction. In the case study, we attempted where possible to make the direction of implication arcs within the Bayesian Networks consistent with the top-down decomposition found in the existing software specifications.

## 2.1 An Example of a Bayesian Network Implementation of SRWs

Figure 1 models a SRW fragment as a Bayesian Network. This example models the situation where a requirement for PDES (Parallel Discrete Event Simulation) and a requirement for Distributed_Sim (Distributed Simulation) imply a requirement for time_mgmt (time management). The node time_mgmt_msgs (time management messages) is implied by time_mgmt.

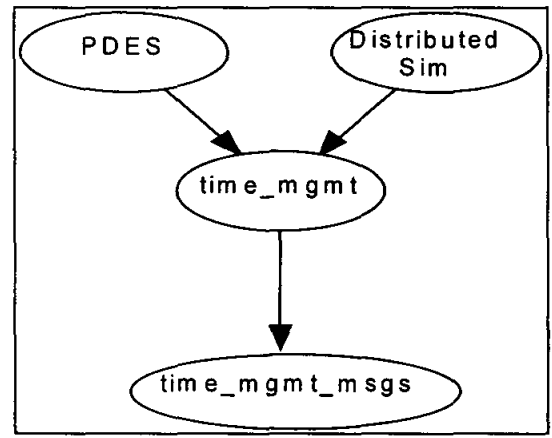

Figure 1: SRW as a Bayes Net

As with most requirements analysis tasks, the example assumes that there was little enumerative data to base a frequentist assessment of conditional probabilities. Therefore, the determination for the conditional probability tables was done heuristically. First, probabilities were selected to model that fact that both PDES and Distributed_Sim are, with no other evidence, very unlikely. Setting the initial value that they are implied to 0.2 and the value that they are not implied to 0.8 reflected this.

Next, conditional probabilities were selected to model the situation where it is likely that time_mgmt is implied if both PDES and Distributed_Sim are implied. This was accomplished by setting the conditional probability to 0.8 if both PDES and Distributed_Sim are implied and to 0.2 if neither one is implied. To provide a modicum of support if only one of the source nodes is implied, the probability that time_mgmt is implied if either PDES or Distributed_Sim is implied was set to 0.4.

The last conditional probability table that was created was for time_mgmt_msgs. The modeling criteria chosen was that unless time_mgmt is implied it was unlikely that time_mgmt_messages would be implied. This was reflected by defining the conditional probability that time_mgmt_msgs was implied as 0.2 if time_mgmt was not implied and 0.8 if time_mgmt_msgs was implied. These probability assessments are summarized as follows:

- P(pdes) = P(distributed_sim) = implied = 0.2.

- P(time_mgmt) = implied

|  | Distributed_Sim Implied | Distributed_Sim Not implied |
|---|---|---|
| PDES Implied | (0.8) | (0.4) |
| PDES Not implied | (0.4) | (0.2) |

- P(time_mgmt_msgs) = implied

| time_mgmt_implied | time_mgmt_not implied |
|---|---|
| (0.8) | (0.2) |

With no evidence introduced, the marginal probability that the time_mgmt is implied is implied is 0.28 and that time_mgmt_msgs is implied is 0.35. This is interpreted that in the absence of any evidence the requirements time_mgmt and time_mgmt_msgs are probably not appropriate.

Suppose that concrete evidence was received that Distributed_Sim is definitely needed. Since this is now a certainty, the Distributed_Sim node can be considered an observed fact, which has the effect of setting the implied probability to 1.0 and the not implied probability to 0. This belief then propagates through the network raising the confidence that time_mgmt is implied to 48% and time_mgmt_msgs to 49%. Further, if the system requirement PDES is known the probability is raised to 80% for time_mgmt and 68% for time_mgmt_msgs.

For the purpose of reporting both to the user and for the consumption of the agents, we created a threshold probability for requirements to be declared as implied. In the example, suppose that the threshold is 0.75. Only upon the introduction of both pieces two pieces of evidence would a requirement be declared as implied, where the evidence is defined as Distributed_Sim or PDES. Specifically, the requirement for time_mgmt would be declared as implied and time_mgmt_msgs would not as P(time_mgmt_msgs=implied) = 0.68. If there is other knowledge to indicate that time_mgmt_msgs should have been implied, it may be appropriate to modify the model or alter the threshold.

The example described here was kept purposefully small. In practice, each SRW fragment contained between 30 and 60 interconnected system requirements represented as random variables in a Bayesian Network. Figure 2 shows one such fragment from the case study, time2.xml, which modeled requisite time management functionality. When SRW fragments were glued together, the resultant SRWs typically had about 200 nodes.

## 3 System Overview

The BOSH architecture was designed to demonstrate the use of SRWs in the translation of user requirements to system requirements. BOSH was inspired by Joint

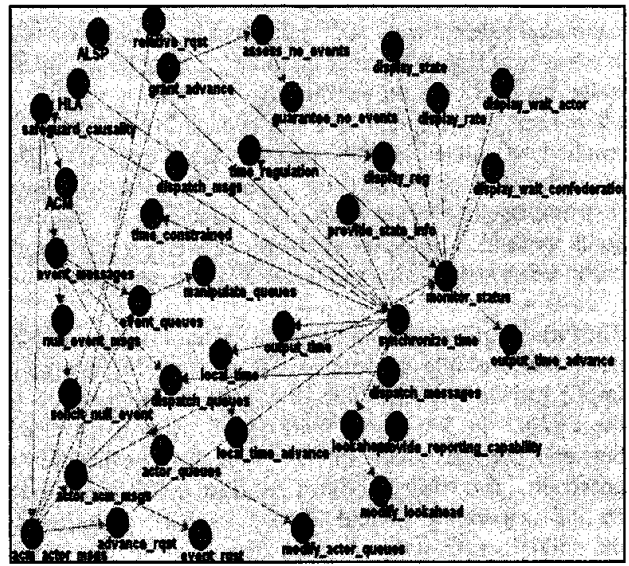

Figure 2:  Example SRW Fragment:  time2.xml

Requirements Planning Sessions where a group of experts will collaborate on defining system requirements. BOSH uses an architecture in which software agents mimic the experts and the SRW represents the experts' joint knowledge of the relationships between requirements. Conceptually the SRW can be thought of as a blackboard, a common knowledge structure for the dissemination of information between agents. A top-level view of BOSH is provided in Figure 3.

SRW Agents make determinations about whether a node is likely to be implied or not implied. To do this, they use information created by other agents as well as the results of weak implication within the SRW. This information is communicated between agents using a common agent communication language. The effect of various actions such as the implication and not implication of nodes in the SRW is observed directly by the SRW Agent. SRW Agents can also elect to assert that a node in the SRW is implied, or retract the implication of a node that has been implied. When this happens, all other agents are notified of the assertion. We also allow agents to declare soft evidence representing probable implication.

Conflicts between agents are not explicitly modeled and the resolution is left to the operator. It is possible that this lack of conflict modeling could result in a race condition between agents sequentially reversing each others implication actions, but this has not been observed in practice.

The BOSH architecture has been implemented in Java. BOSH uses a Bayesian Network representation of SRWs implemented using a freeware application programmers interface (API), JavaBayes v0.33 [COZ98]. Individual SRWs are glued together using a tool written for specifically for this research. The translator from natural language user requirements to the system's internal representation is based on a Java version of the ELIZA program [WEI66] and the agents are production rule systems based on JESS.

To assess the effect of the introduction of new evidence, the algorithm propagated evidence in the glued Bayesian network every time a message was received. This resulted in significant computation time when the number of agents exceeded eight and the number of nodes within the SRW exceeded 250. While BOSH is a prototype implementation, using BOSH on a relatively modest computer such as a Pentium II running at 266 MHz with 64 Mbytes of RAM allows little room for expansion. It should be noted, however, that BOSH is written entirely in JAVA and at the time of this writing, Java was significantly slower than other languages such as C++. Performance enhancements could certainly be achieved with recoding.

## 4 Experimental Results

SRWs were used in the generation of a technical specification for a distributed simulation management tool. While this tool is required to perform much of the same functionality as an existing tool, it must interface with completely different infrastructure software as well as provide a significant amount of new functionality. The problem was representative of the target application for SRWs; existing documented software artifacts made domain decomposition and SRW generation relatively straightforward. User requirements were also well defined and documented eliminating the need for an elicitation exercise.

The task began by a decomposition of the user requirements into five large functional areas: interaction between components, communications considerations, display of simulation objects, save functionality and time management. Concurrent with this exercise a population of 23 SRWs and 23 SRW agents was generated from the software artifacts. Each agent and was designed to model a particular viewpoint into the requirements space which roughly coincided the decomposition previously documented in the existing software artifacts. For example, SRWs were constructed for crash recovery, attribute visibility, screen management, user interface issues, infrastructure management and time management as the domain being modeled was that of parallel discrete event simulation.

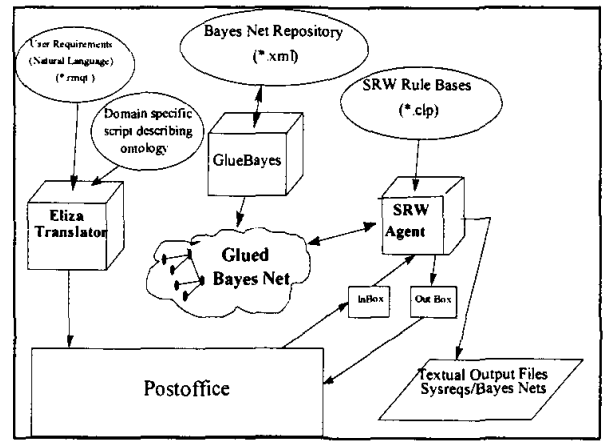

Figure 3: BOSH Architecture

We then identified the SRWs that had system requirements relevant to each of the functional areas. These SRWs were glued together to form five larger SRWs. From the 23 previously defined agents, we then defined sub-populations of SRW agents that contained knowledge about the broad functional areas. Members of these sub-populations would be able to interpret how user requirements engender specific system requirements in the glued SRW.

User requirements were sent to the agents as messages. As appropriate, the agents interpreted the user requirements as hard or soft evidence for system requirements within the SRW. As the evidence was declared, belief propagated through the SRW. System requirements that exceed the threshold were announced as implied system requirements. These implied system requirements were saved to form the core of the technical specification.

Validation of the results came from expert evaluation. After the initial runs, the system requirements were repackaged into a technical description document that was submitted to software engineers working on the development of the tool. The experts provided an evaluation and this was used to modify the SRWs and the SRW agents. After modifications were made, a second iteration was run with the revised agents and SRWs. This was then submitted to the technical experts a second time for their evaluation and comment.

The first two phases of testing were run in expert mode. Expert mode is defined by setting a relatively low threshold (P(implied) > threshold) for a system requirement to be declared as implied. This resulted in most of the requirements that should be implied by the system implied, but also resulted in a number of incorrectly implied requirements. The thinking behind the expert mode is that an expert user can discard the obviously incorrect requirements in favor of getting the most amount of coverage. This is precisely what happened in the first two phases of the case study. Since we could be considered experts in the domain of interest, the threshold was set purposely low and requirements that were obviously incorrect were thrown out.

During the assessments of the generated document, the evaluation team was unaware that BOSH was being used to generate the requirements. This was done purposefully so as not to introduce either a positive or a negative bias into the experiment. This is also why we ran the system in expert mode. Obvious errors in the document would have alerted the evaluation team and led to bias in the evaluation.

To assess how well the system would perform with no without eliminating obviously incorrect answers, a naïve user test was run. The naïve user simply regurgitated the results from BOSH with no additional processing. This test was run using the SRWs that were tuned from phase one. The results were evaluated against the corrections made to the phase two document. A summary of this evaluation is provided in Figure 4.

Requirements were evaluated according to two broad criteria, accuracy and coverage. Accuracy refers to whether the system requirement was correctly implied and was further broken down into the categories of correct as written, partially correct or miscategorized. Coverage referred to whether a requirement that should have been implied was implied.

For phase one of the test approximately 70% of the requirements BOSH indicated should be implied were correct as written, 13% were partially correct and 17% were incorrect. BOSH implied 85% of the requirements that should have been implied. After tuning, the user requirements were run again and significant improvement was realized. Specifically, 91% of the requirements that BOSH implied were implied correctly, 5% were partially correct and 4% were incorrect. BOSH implied 90% of the requirements that should have been implied. For the naïve user test, the results were approximately equal to those realized during phase one of the test. BOSH correctly implied 65% of the system requirements correctly, implied 29% partially correctly and 6% were incorrect. BOSH achieved an 82% coverage ratio during the naïve user test.

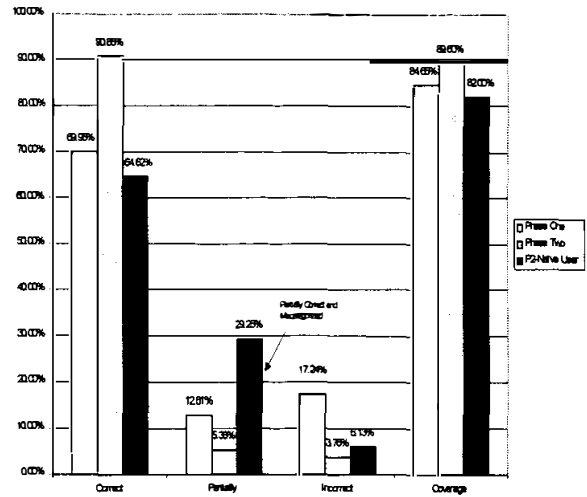

Figure 4: Summary Test Results

While providing anecdotal evidence, several observations can be made from these tests. First, significant improvements are possible with tuning. This would indicate that as the population of SRWs is used, the performance can rapidly increase to the point that useful information is readily obtained. Secondly, the naïve user mode demonstrated that a population of SRWs can become useful to even novice users after only one iteration of modifications. In this case, with minimal guidance the naïve user can roughly approximate the performance of an expert using the system on the first iteration. While anecdotal, this data indicates that an evolving system of SRWs can provide even novice users with a quick snapshot of the necessary system requirements based upon user input. As would be expected, expert users get better results but that is consistent with most decision support tools as shown in [LEH93].

This case study is also interesting in that the tool for which the specification is being created is actively under development. Although there is a significant amount of functionality already implemented no design documentation had been developed. Verification of requirements became an exercise of determining if functionality had in fact been implemented in the tool. The SRW population in some cases suggested requirements that were not implemented due to design decisions. This identification of missed requirements was an unanticipated benefit of the approach.

## 5 Related Work

The authors have not found any research that has specifically married agentry with Bayesian reasoning to address the problem of translating user requirements to system requirements. However, a number of other efforts have pursued promising research avenues. For example, Rich [RIC87] discussed the creation of a requirements apprentice that assists in the creation and modification of software requirements. Rich's work focused on the translation of informal requirements to formal specifications through an extensive model of the domain. This is somewhat similar to Reubenstein's Requirements Apprentice [REU91] that uses techniques such as dependency-directed reasoning and reuse of clichés to transition informal requirements to formal statements of need. Krause also discusses automated requirements engineering support in [KRA97].

Maiden [MAI95] has conducted work to define a set of formal problem abstraction as part of a large European effort in requirements engineering entitled NATURE (Novel Approaches To Requirements Engineering). Maiden's approach is derived from a belief that most software engineering problems belong to a tractable set of hierarchical classes. Towards this end approximately 150 classes have been derived from software engineering problems, domain analysis and textbooks.

The concept of a number of agents cooperating to develop a requirement specification has been suggested as part of the viewpoints research (e.g., [FIN92], [EAS94], [NUS94]). Viewpoints came out of the realization that the requirements for many systems are elicited from multiple perspectives. These perspectives can overlap, contradict one another or complement each other.

Liu [LIU98] has conducted research using Quality Function Deployment techniques with fuzzy logic to assess imprecise requirements. By putting a fuzzy front end on a QFD process, Liu asserts that one can better understand the true meaning of user requirements that would provide for a more effective mapping of user requirements to system requirements.

## 6 Summary and Future Work

This paper presented a methodology for representing the relationships between system requirements within a domain. An abstract concept called the System Requirement Web was introduced which was defined as a directed graph whose nodes are connected by the weak implication operator. SRWs have been instantiated as Bayesian Networks. A tool has been created which has been used to generate a technical specification using the Bayesian Network implementation of SRWs. Good results were achieved after minimal tuning.

The case study discussed in this paper used SRWs on a problem that fit most of the criteria for successful SRW usage. In particular, the domain was well defined with a number of software artifacts in existence. The technology being used while novel in an engineering sense was well established. Further, experts were available to both assess the output of BOSH as well as assist in the construction of SRWs. These conditions suggest that the best of use of SRWs is for targeted applications, as the knowledge engineering in developing the SRWs is non-trivial. Further, the reusability of SRWs is domain dependent with the specificity of the SRW being inversely related to its general applicability.

Representing SRWs as Bayesian networks was done in this research from primarily a toolsmith approach, meaning that the Bayesian networks were used essentially as black box implementations. While favorable results were obtained using heuristic simplifications to assess prior and conditional probabilities, it is intriguing to wonder if a finer grained representation of the SRWs would produce a better model of the system requirement space. Particularly, a homogeneous approach to assessing conditional probability tables while expedient is not the most sensitive way to model the domain. It is likely that a more refined rule based approach to determining conditional probabilities within the SRW while preserving the ease of construction would offer higher fidelity in modeling implication.

Similarly, a more descriptive scheme for assessing the quality of output would be useful. All system requirements modeled in the SRW are not equally important to the system. Weak implication models how much a requirement is needed; there was no explicit modeling as to how important that requirement was. In pilot studies the SRW node description was expanded to include a context sensitive importance factor but this was not fully explored.

Another opportunity for further research is in the dynamic tailoring of SRWs predicated on some input criteria. SRWs are currently reasonably static structures, modifiable upon user intervention but only tailorable from the perspective of the propagation of belief and the creation of dynamic evidence nodes. However, it would be interesting to explore tailoring SRWs dynamically when they are about to be. This tailoring could take the form of deleting nodes that are clearly not of interest to the SRW agents which will analyze the SRW. In addition to performance increases within the prototype, a cognitive simplification could result, as networks of several hundred nodes are virtually impossible to decipher.

The SRW approach to developing technical specifications has been developed and demonstrated. The use of Bayesian Networks was shown to be an effective method for implementing an abstract concept for representing the relationships between requirements. Future research will continue to provide an evaluation as to the suitability of modeling uncertainty in requirements engineering.